\documentclass[final,5p,times,twocolumn]{elsarticle}

\usepackage[hidelinks]{hyperref}
\usepackage{graphicx}
\usepackage{amsmath}  
\usepackage{xspace}

\begin{document}

\begin{frontmatter}

%% Note: \pmbanner before the actual title
\title{Organic electronic-based neutron detectors}

\author[1]{Adrian J. Bevan\corref{cor1}%
  \fnref{fn1}}
\ead{a.j.bevan@qmul.ac.uk}

\author[1]{Fani E. Taifakou}

\author[1]{Choudhry Z. Amjad}

\author[1]{Aled Horner}

\author[2]{C. Allwork}

\author[1]{A. J. Drew}

 \cortext[cor1]{Corresponding author}

\affiliation[1]{organization={Queen Mary University of London}, 
                 addressline={School of Physical and Chemical Sciences, Mile End Road},
                 postcode={E1 4NS}, 
                 city={London}, 
                 country={UK}}
\affiliation[2]{organization={Atomic Weapons Establishment}, 
                 addressline={Aldermaston},
                 postcode={RG7 4PR}, 
                 city={Reading}, 
                 country={UK}}

\begin{abstract}
  In recent decades organic electronics has entered the mainstream of consumer electronics, driven by innovations in scalability and low
  power applications, and low-cost fabrication methods. The potential for using organic semiconductor electronic devices as radiation
  detectors, and in particular for neutron detection is reported. We report results of laboratory tests using $\alpha$ particles as well
  as the response to thermal and fast neutrons covering the energy range 0.025 eV to 16.5 MeV. GEANT4 simulations are used to provide a
  detailed understanding of the performance and potential of this emerging technology for radiation detection.
\end{abstract}


\end{frontmatter}

%text of the article

%% Use \section commands to start a section
\section{Introduction}
\label{sec:intro}
%% Labels are used to cross-reference an item using \ref command.
Organic semiconductors are capable of detecting highly ionising radiation via both $\alpha$ particle detection as well as protons~\cite{Beckerle:2000,Taifakou:2021,Fratelli:2021}.  Results have also been published on fast~\cite{Kargar:2011} and thermal neutron detection~\cite{Chatzipiroglou:2020}, however the thermal neutron detection reported in~\cite{Chatzipiroglou:2020} neglects fast neutron contamination of the source used. Here we discuss recent results from our group on the detection of fast and thermal neutrons. The fast neutron interactions are via elastic or inelastic scattering with the organic semiconductor material (depending on the neutron energy), and thermal neutron detection is facilitaed via boron neutron capture (BNC).  We use isotopically enriched boron carbide (B$_4$C), with a $^{10}$B fraction of 96.6\% by weight.  

The ultimate aim of these developments is to be able to make a cheap and scalable neutron detector that can be used for radioactive material safeguarding, however neutron detectors are used in a range of commercial, government and academic applications.  We present results on fast and thermal neutron detection in these proceedings, summarising 
the work presented in~\cite{Borowiec:2022krn}, and ancillary related studies.  We are exploring possible applications of the current technology and working to understand the limitations as noted below.

\section{Devices and Radiation Sources}
\label{sec:devices}

We fabricate $\pi$-conjugated organic semiconductor radiation detectors following the approach detailed in ~\cite{Borowiec:2022krn}.  We have explored the performance of devices fabricated using several polymers including 
poly{[N,N$\,^\prime$-bis(2-hexyldodecyl)naphthalene-1,4,5,8-bis(dicarboximide)-2,6-diyl]-alt-5,5$\,^\prime$-(2,2$\,^\prime$-bithiophene)} (PNDI(2HD)2T).  
We refer to this as PNDI here.   Further work on related polymers, an octyldodecyl variant of PNDI and carborane containing napthalene polymers (paper in preparation). Results reported here are on diode-like structures, however we also working with field effect transistors (FETs) and combinations of diodes and FETs (paper in preparation). Devices are fabricated on a range of substrates including soda lime glass, high and low density polyetheleyne.  The devices have a thickness of 30-40$\mu$m, and each substrate has four 2 mm$\times$2 mm pixels.  As noted above for devices targeting measurement of thermal neutrons we deliberately contaminate the polymer with B$_4$C. We choose this over boron for two reasons. Firstly boron, as a metal, could create shorts across the electrodes unless coated in an insulator.  Secondly the energy levels of the B$_4$C fall outside of the band gap of the polymers that we use. This means that the introduction of the contaminant (to first order) does not introduce traps or significantly change the electronic properties of our detectors.

%\section{Radiation sources}
%\label{sec:sources}

We used the neutron facility at the UK's National Physical Laboratory (NPL) in Teddington to test our technology.  We have used the thermal pile~\cite{Hawkes:2018}. Monochromatic fast neutrons are generated via $\,^7$Li(p,n)$\,^7$Be for the 0.565 MeV beam and via T(d,n)$\,^4$He for the 16.5 MeV beam data. 

\section{Results}
\label{sec:results}

\begin{figure}[!ht]%% placement specifier
\centering%% For centre alignment of image.
\includegraphics[width=9.0cm]{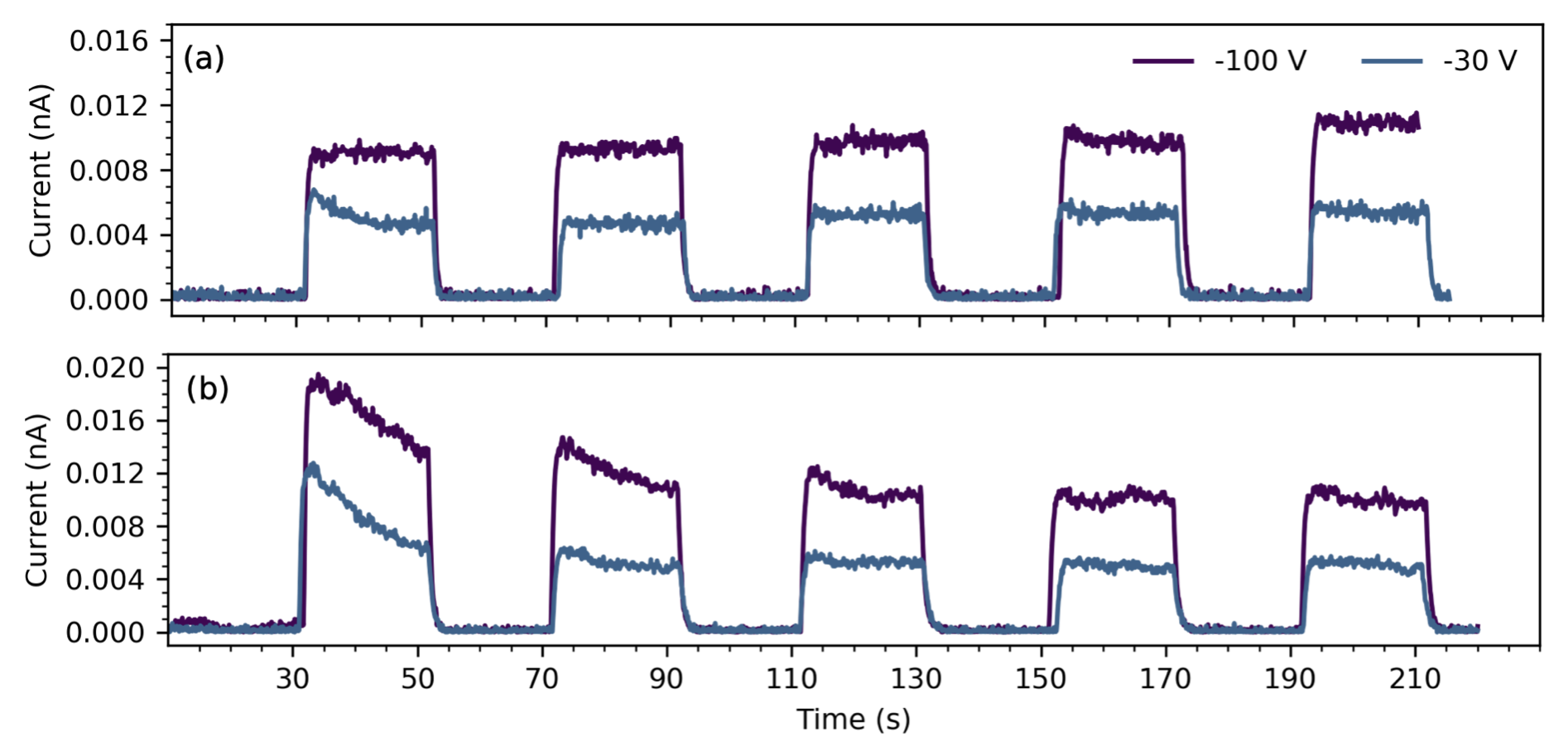}
\caption{The current enhancement observed on exposing the devices to fast neutrons: (top) 0.565 MeV, (bottom) 16.5 MeV.}\label{fig1}
%% https://en.wikibooks.org/wiki/LaTeX/Importing_Graphics#Importing_external_graphics
\end{figure}

We present the results of exposing our devices to fast neutrons in Figure~\ref{fig1}.  We observe signals as an increase above the dark current when exposing to radiation.  Similar results are obtained with an AmBe source (see~\cite{Borowiec:2022krn}).  The signal variations are from beam current variations~\cite{Borowiec:2022krn} arising from the response time from the accelerator feedback system.

\begin{figure}[!ht]%% placement specifier
\centering%% For centre alignment of image.
\includegraphics[width=9.0cm]{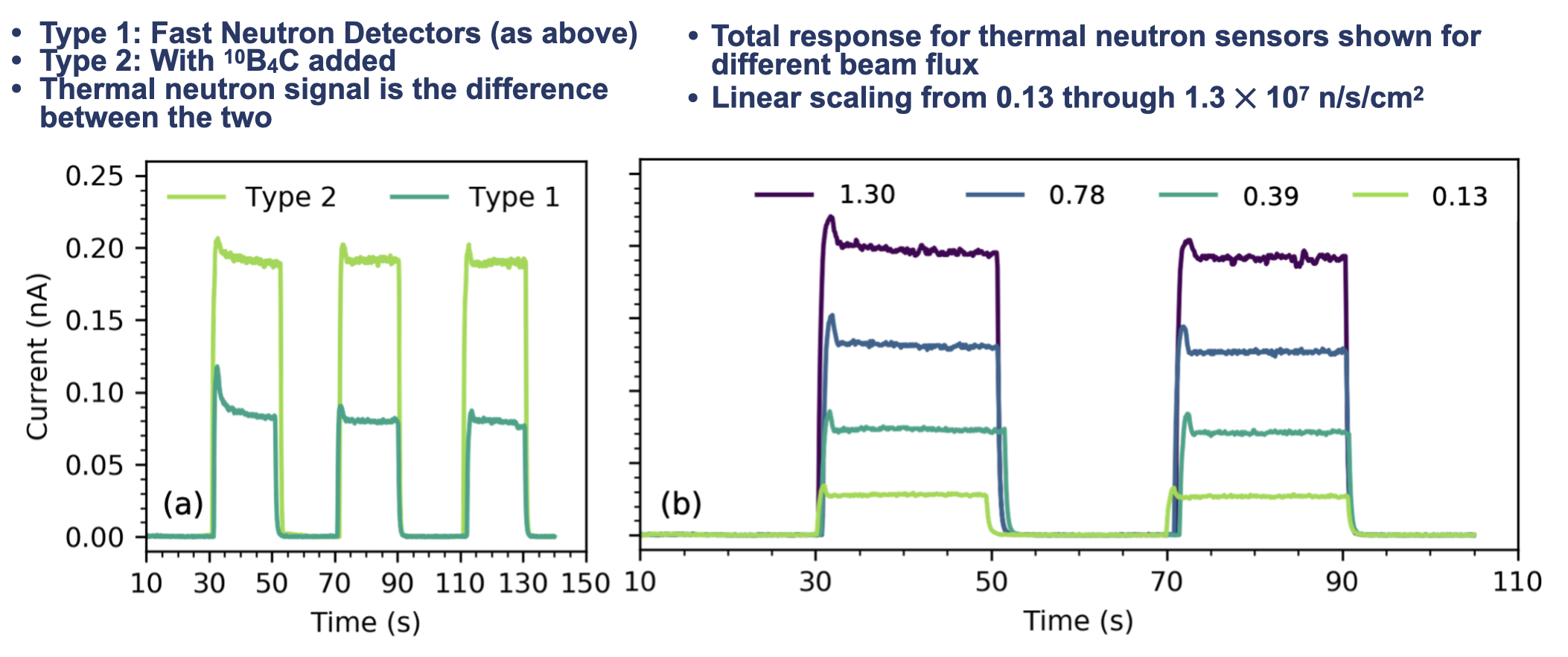}
\caption{The current enhancement observed on exposing the devices to neutrons in the thermal pile. (left) the difference in response between devices with and without B$_4$C, corresponding to thermal neutrons. (right) the device response (with B$_4$C) for different thermal neutron fluxes.}\label{fig2}
%% https://en.wikibooks.org/wiki/LaTeX/Importing_Graphics#Importing_external_graphics
\end{figure}

The thermal pile at NPL is a source of predominantly thermal neutrons, with two measurement points.  The one we use is a narrow bore access hole that enables measurement with fluxes up to $1.3\times 10^7$ $n cm^{-2}/s$.  The exact form of the fast neutron background in this location of the thermal pile has not been measured, however it is assumed to be similar to the thermal column measurement point that has been reported in~\cite{Kolkowski:1999}.  As a result we perform measurements of devices with and without B$_4$C. This allows us to compute the difference in two otherwise equivalent devices to determine the signal response from thermal neutrons and disentangle the fast neutron background component.  The thermal and fast neutron response of our devices in the thermal pile are shown in Figure~\ref{fig2}. The photon detection efficiency is low for these thin, low $Z$ detectors. The gamma response from a 3.7MBq $^{60}$Co source is not detectable. 

\section{Summary and future work}
\label{sec:summary}

We have tested organic semiconductor radiation detectors tailored for fast and thermal neutron detection.  These have been fabricated using drop casting and chemistry compatible with scalable technologies used in industry.  We have demonstrated fast neutrons detection for energies up to 16.5 MeV, and illustrated how to disambiguate fast and thermal neutron contributions from data in facilities such as the NPL's Thermal Pile.

We are exploring applications of this technology for beam monitoring systems, where we are currently designing a fast neutron beam telescope. The initial prototype will consist of a pair of 5$\times$5 cm devices, one to determine the fast neutron component of the signal and the second to determine the thermal neutron component of the fast neutron signal.  Measurements from these tests are intended to inform the segmentation of a multi layer pixelated `neutron camera' system.  We have evaluated the technology up to a fluence of $4\times 10^{10} n/cm^2$ over a period of a few hours, the majority of that being delivered in an extended run in NPL's thermal pile over an hour.  A detailed study of the radiation tolerance of this technology is needed to understand the limits of the technology, and we are planning to study this the ISIS neutron beam facility in the UK.

%% The Appendices part is started with the command \appendix;
%% appendix sections are then done as normal sections
\section*{Acknowledgements}

We dedicate these proceedings to the memory of our colleague and friend, Dr Theo Kreouzis, who sadly passed away toward the end of 2023. Theo's contribution to our work helped us achieve the results presented here, and continues to inspire our work with this technology.

\noindent
{\bf Copyright 2024 UK Ministry of Defence © Crown Owned Copyright 2024/AWE}

  \bibliographystyle{elsarticle-num-names} 
  \bibliography{bibfileBevanPisaMeetingProcsOrganics}

\end{document}